# A Deep Learning-Based GPR Forward Solver for Predicting B-Scans of Subsurface Objects

Qiqi Dai, Yee Hui Lee, *Senior Member, IEEE*, Hai-Han Sun, Jiwei Qian, Genevieve Ow, Mohamed Lokman Mohd Yusof, and Abdulkadir C. Yucel, *Senior Member, IEEE*

*Abstract*—The forward full-wave modeling of ground-penetrating radar (GPR) facilitates the understanding and interpretation of GPR data. Traditional forward solvers require excessive computational resources, especially when their repetitive executions are needed in signal processing and/or machine learning algorithms for GPR data inversion. To alleviate the computational burden, a deep learning-based 2D GPR forward solver is proposed to predict the GPR B-scans of subsurface objects buried in the heterogeneous soil. The proposed solver is constructed as a bimodal encoder-decoder neural network. Two encoders followed by an adaptive feature fusion module are designed to extract informative features from the subsurface permittivity and conductivity maps. The decoder subsequently constructs the B-scans from the fused feature representations. To enhance the network's generalization capability, transfer learning is employed to fine-tune the network for new scenarios vastly different from those in training set. Numerical results show that the proposed solver achieves a mean relative error of 1.28%. For predicting the B-scan of one subsurface object, the proposed solver requires 12 milliseconds, which is 22,500x less than the time required by a classical physics-based solver.

## I. Introduction

GROUND-PENETRATING radar (GPR) is a non-destructive technique that uses electromagnetic (EM) waves to inspect subsurface environments. The forward full-wave modeling of the GPR system is of great significance for understanding the subsurface scattering mechanisms and interpretation and inversion of the GPR data [1]. Many forward full-wave solvers have been developed so far to characterize the radar signatures of subsurface objects. These solvers are primarily based on the finite-difference time-domain (FDTD) method [2], the method of moments [3], and the finite element time-domain method [4]. However, these physics-based traditional solvers require excessive computational time, especially when their repetitive executions are required by signal processing or machine learning-based inversion algorithms. In particular, full-waveform inversion (FWI) algorithm requires the execution of a forward GPR solver in every iteration of its optimization stage, and an inversion of each GPR data often requires hundreds or even thousands of iterations [5]. Furthermore, the machine learning-based algorithms developed to solve GPR inverse problems [6] require a large set of data for their training and testing stages. For example, over 400,000 GPR B-scans are used to train the deep convolutional neural network for imaging subsurface scenarios [7]. Using a traditional physics-based forward solver, the dataset generation becomes highly time-consuming.

To alleviate the computational burden of traditional GPR forward solvers, machine learning-based fast forward solvers have been proposed [8], [9]. For example, the work in [8] explores the relationship between model parameters and first arrival travel time, while the technique in [9] forms the relationship between the properties of the concrete and rebars and their corresponding 1D A-scans. All these techniques are proposed to form the parameter-to-parameter relationship between the input parameters related to the GPR scenario and the output parameters, which only yield partial information about the GPR scenario. For example, the A-scan in [9] only depicts the reflected signal at one position, which only contains information in the narrow spatial domain that affects the signal. It cannot capture the reflected signals at different positions that contain more object information. Furthermore, the applied subsurface scenarios are simple and limited in the training domain. So far, there exists no study exploiting deep learning algorithms for GPR forward modeling and forming a mapping relationship between the input image of the GPR scenario and the output image (e.g., B-scan) that can provide the complete 2D information of the subsurface scenario.

In this study, a deep learning-based GPR forward solver for characterizing 2D radar signatures of subsurface convex objects is proposed. A deep encoder-decoder neural network is designed to predict the GPR B-scan for given permittivity and conductivity maps of the subsurface scenario. The bimodal encoder of the proposed network independently extracts informative features from these two input maps. To enhance the cross-modal connectivity, a feature fusion module with cross attention is employed to adaptively fuse the extracted bimodal feature representations. The fused feature representations are then fed into the decoder to produce the B-scan. The network is

This work was supported by the Ministry of National Development Research Fund, National Parks Board, Singapore (*Corresponding authors: Yee Hui Lee; Abdulkadir C. Yucel*).

Q. Dai, Y. H. Lee, H. -H. Sun, J. Qian, and A. C. Yucel are with the School of Electrical and Electronic Engineering, Nanyang Technological University, Singapore 639798 (e-mails: qiqi.dai@ntu.edu.sg, eyhlee@ntu.edu.sg, haihan.sun@ntu.edu.sg, qian0069@e.ntu.edu.sg, acyucel@ntu.edu.sg).

G. Ow and M. L. M. Yusof are with the National Parks Board, Singapore 259569 (e-mails: genevieve_ow@nparks.gov.sg, mohamed_lokman_mohd_yusof@nparks.gov.sg).

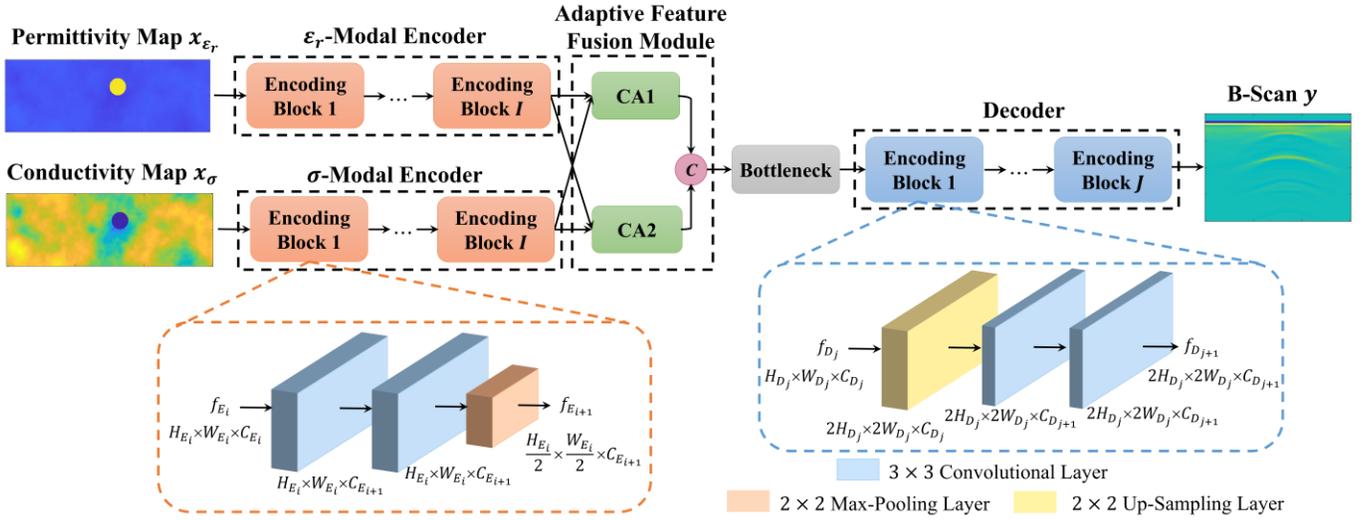

Fig. 1. The structure of the proposed network. $i, j$ and $I, J$ are the indices and total numbers of encoding and decoding blocks, respectively. "$c$" in the pink circle represents the operation of concatenation. $f_{E_i}$ and $f_{D_j}$ represent the feature maps of the $i^{th}$ encoding block and the $j^{th}$ decoding block in the encoder and decoder, respectively. $H_{E_i}, W_{E_i}, C_{E_i}$ and $H_{D_j}, W_{D_j}, C_{D_j}$ are the heights, widths, and channel numbers in the encoder and decoder, respectively.

trained with a diverse set of subaurface permittivity and conductivity maps and their corresponding B-scans under heterogeneous soil conditions. Transfer learning is introduced to boost the network' generalizability for new scenarios. The test results show that the proposed solver using the mean square error (MSE) loss for training achieves the lowest mean relative error (MRE) of 1.28% and a near-real-time B-scan prediction in 12 milliseconds.

## II. METHODOLOGY

Let $x$ and $y$ denote a subsurface scenario and its corresponding B-scan obtained via GPR scanning. A deep encoder-decoder neural network is proposed to form the mapping $h(\cdot)$ from $x$ to $y$. The B-scans obtained from an FDTD solver are used to train the network. Using the well-trained network, the B-scans can be automatically and rapidly predicted from given subsurface scenarios via $y = h(x)$. The relative permittivity ($\varepsilon_r$) map $x_{\varepsilon_r}$ and conductivity ($\sigma$) map $x_\sigma$ of the subsurface scenario $x$ are set as the inputs of the network, as shown in Fig. 1. The output of the network is the corresponding B-scan via $y = h(x_{\varepsilon_r}, x_\sigma)$. The network structure and learning method are described as follows.

### A. Bimodal Encoder

As shown in Fig. 1, the left path of the network includes a $\varepsilon_r$-modal encoder and a $\sigma$-modal encoder, which independently extract informative features from the input permittivity map $x_{\varepsilon_r}$ and conductivity map $x_\sigma$. Each encoder consists of $I$ consecutive encoding blocks. Every encoding block is made up of two consecutive $3 \times 3$ convolutional layers with the strides of $1 \times 1$, followed by a $2 \times 2$ max-pooling layer with the stride of $2 \times 2$ for down-sampling. The rectified linear unit (ReLU) activation follows after every convolutional layer to provide high non-linearity of the network. Assume the dimension of the input feature map $f_{E_i}$ in the $i^{th}$ encoding block is $H_{E_i} \times W_{E_i} \times C_{E_i}$, the output feature map $f_{E_{i+1}}$ becomes $\frac{H_E}{2} \times \frac{W_E}{2} \times C_{E_{i+1}}$.

Finally, a $\varepsilon_r$-modal feature representation $F_{\varepsilon_r}$ and a $\sigma$-modal feature representation $F_\sigma$ are obtained from the two encoders.

### B. Adaptive Feature Fusion Module

To enhance the connectivity between $F_{\varepsilon_r}$ and $F_\sigma$ and capture cross-modal information, an adaptive feature fusion module is designed [Fig. 2]. In this module [Fig. 2], the inputs of a self-attention block [10] are the vectors of query $Q$, key $K$, and value $V$, projected by a linear feed-forward layer (the $1 \times 1$ convolution). The output attention feature $A$ is produced by weighted summation over $V$ as formulated in Eq. (1). The dot products of $Q$ and $K$ are computed, scaled by $\sqrt{d}$ ($d$ is the dimension of $K$), and followed by a Softmax function to obtain the weights on $V$.

$$A(Q, K, V) = Softmax\left(\frac{Q \cdot K^T}{\sqrt{d}}\right) \cdot V \quad (1)$$

To adaptively fuse the bimodal features in $F_{\varepsilon_r}$ and $F_\sigma$, two cross-attention (CA) blocks [11], CA1 and CA2, are employed. The query of $F_{\varepsilon_r}$ ($Q_{\varepsilon_r}$) and that of $F_\sigma$ ($Q_\sigma$) are exchanged with each other, which introduces cross-modal interactions between the $\varepsilon_r$-modal and $\sigma$-modal features. The $A_{\varepsilon_r}$ in CA1, which represents the $\varepsilon_r$-modal attention features obtained with the guidance of $\sigma$-modal information, and the $A_\sigma$ in CA2, which represents the $\sigma$-modal attention features obtained with the guidance of $\varepsilon_r$-modal information, are computed by

$$A_{\varepsilon_r}(Q_\sigma, K_{\varepsilon_r}, V_{\varepsilon_r}) = Softmax\left(\frac{Q_\sigma \cdot K_{\varepsilon_r}^T}{\sqrt{d}}\right) \cdot V_{\varepsilon_r}, \quad (2)$$

$$A_\sigma(Q_{\varepsilon_r}, K_\sigma, V_\sigma) = Softmax\left(\frac{Q_{\varepsilon_r} \cdot K_\sigma^T}{\sqrt{d}}\right) \cdot V_\sigma. \quad (3)$$

To jointly learn the information from different representation subspaces, multi-head attention [10]-[11] structure is adopted in both CA1 and CA2. The functions (2) and (3) are performed on $m$ paralleled heads that are linearly projected from $Q, K$, and $V$. The output features of each head are concatenated and linearly projected to $A$, which is once again followed by $1 \times 1$ convolution. Residual connection and layer normalization are applied in each sub-layer. Then two extracted feature representations with cross-modal information are concatenated as the fused feature representation $F_{\varepsilon_r \leftrightarrow \sigma}$.

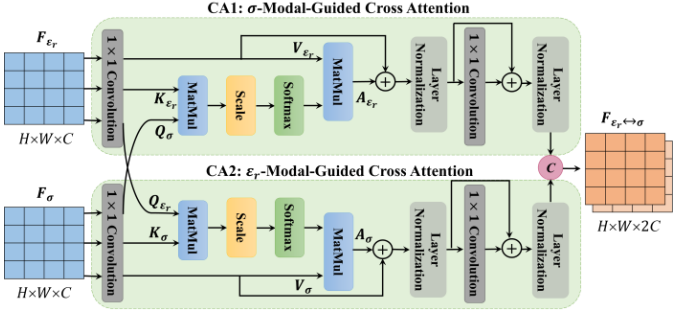

Fig. 2. The structure of the adaptive feature fusion module.

### C. Decoder and Loss Function

A bottleneck block consisting of a 3 × 3 convolutional layer and a 3 × 3 transposed convolutional layer follows to bridge the obtained $F_{\varepsilon_r \leftrightarrow \sigma}$ and the decoder. The decoder includes $J$ repeated decoding blocks to predict the B-scan. As shown in Fig. 1, every decoding block has one 2 × 2 up-sampling layer with the stride of 2 × 2 and two consecutive 3 × 3 convolutional layers with the strides of 1 × 1. In the final stage, a 1×1 convolutional layer with the stride of 1 × 1 followed by a linear activation outputs the B-scan $y$. To train the proposed network with fast convergence rate and effectively suppress large errors, the MSE between the B-scan predicted by the proposed network $h_\theta(\cdot)$ and the ground truth B-scan $\hat{y}$ generated by the physics-based solver is adopted as the loss function; here $\theta$ represents the trainable parameters. The objective function is expressed as

$$\min_\theta \left( h_\theta(x_{\varepsilon_r}, x_\sigma) - \hat{y} \right)^2. \quad (4)$$

### D. Transfer Learning for New Scenarios

One limitation of applying deep learning-based data-driven techniques to EM forward modeling problems is their high dependency on the training data. When the testing data is vastly different from the training data, the network becomes ineffective. To tackle this issue, transfer learning [12] is introduced to predict the B-scans of subsurface objects in new scenarios. In the transfer learning technique, first, the proposed network is pre-trained with a large diverse set of data. It is noted that the pre-training only needs to be done once. Second, using the physics-based solver, a small additional set of training data is generated for new scenarios out of the pre-training set. Third, the pre-trained network is set as the initial state and is further optimized based on Eq. (4) using new training data until convergence. The learning rate should be lower than the one in the pre-training process to avoid exploding gradient issues. Finally, the fine-tuned network with enhanced generalization capability can accurately predict the B-scans for new scenarios.

## III. NUMERICAL RESULTS

### A. Dataset Generation and Network Training

To train and test the proposed network, a diverse set of data is generated using an open-source FDTD-based software gprMax [13] executed on a NVIDIA Quadro RTX8000 GPU. In the simulation scenario, as shown in Fig. 3, a 2D domain with dimensions of 1.5×0.5 m$^2$ is discretized by pixels with dimensions of 0.0025×0.0025 m$^2$. The time window is set as 20 ns. A Hertzian dipole (TX) transmitting a Gaussian waveform with a center frequency of 1 GHz is used. A probe (RX) positioned 20 cm away from TX receives the fields. The TX and RX in a common-offset mode move along one straight scanning trace to obtain the B-scan. The scanning step is 2.5 cm. Convex objects with six shapes, including the circle, ellipse, hexagon, pentagon, quadrilateral, and triangle, are considered as subsurface objects. Each object has a random size, position, relative permittivity, and conductivity. The

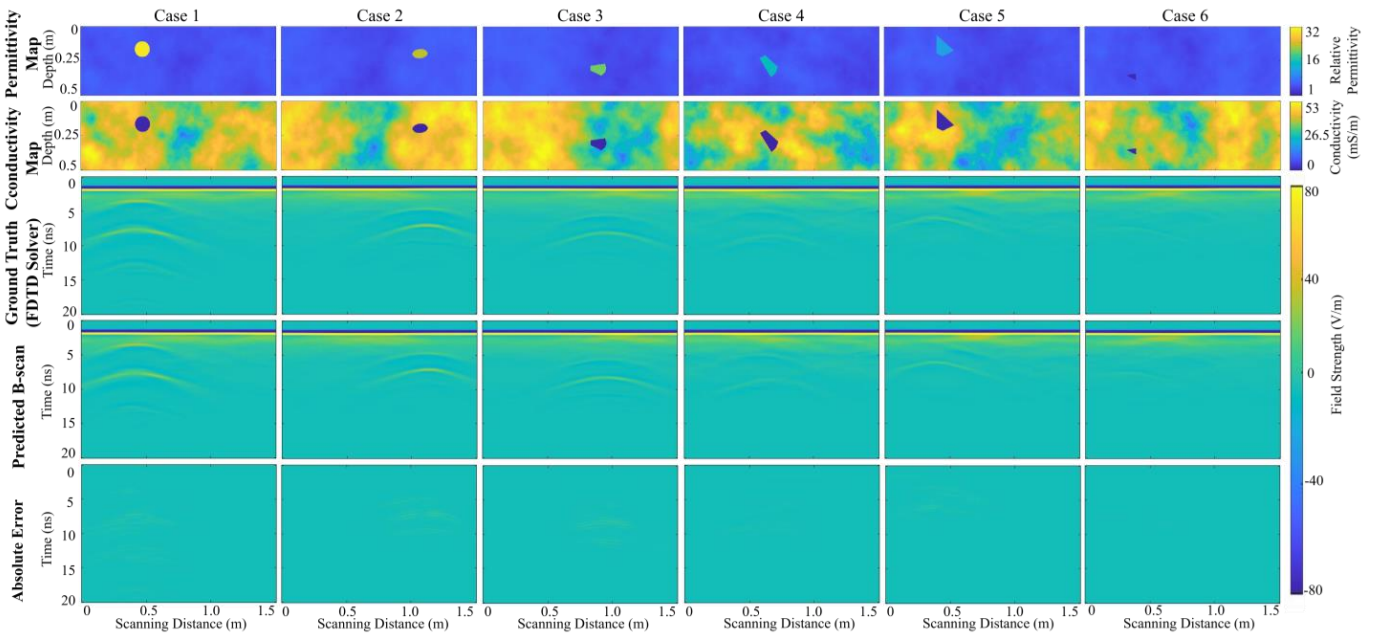

Fig. 3. The comparison of the test results. Ground truth B-scans are obtained by the FDTD solver while the predicted B-scans are the ones obtained using the proposed network. The absolute error figures show the pixel-wise absolute difference between the ground truth and the predicted B-scans.

Peplinski mixing model [14] is used to form the heterogeneous soil environment with realistic dielectric and geometric properties. The soil properties are set as follows: sand fraction 0.5, clay fraction 0.5, bulk density 2 g/cm$^3$, and sand particle density 2.66 g/cm$^3$. 50 different materials over a range of water content from 0.1% to 10%, described by 50 Debye functions, are randomly distributed. The $\varepsilon_r$ and $\sigma$ of the background soil medium vary within [3.59, 7.17] and [8.37, 52.13] mS/m, respectively, while $\varepsilon_r$ and $\sigma$ of the object are randomly selected from [1, 32] and [0, 0.8] mS/m, respectively. Ten random distributions are generated to model 10 diverse heterogeneous soil environments. In total, 15,000 sets of [$x_{\varepsilon_r}$, $x_\sigma$, $\hat{y}$] are generated to train the proposed network, and 1,500 sets are generated to test the performance.

The proposed network implemented on TensorFlow is trained using the obtained dataset. For this purpose, Adam optimizer is used to minimize the loss function (Eq. (4)). Five encoding blocks and five symmetric decoding blocks are employed. The $[C_{E_1}, \ldots, C_{E_5}]$ and $[C_{D_1}, \ldots, C_{D_5}]$ are set as [16, 32, 64, 128, 256] and [256, 128, 64, 32, 16], respectively. $m$ and $d$ are set as 8 and 32, respectively, in the multi-head attention structure. After 100-epoch training, the model with the lowest testing loss predicts $y$ for the given $x_{\varepsilon_r}$ and $x_\sigma$.

### B. The Analysis of Regular Testing Results

To quantitatively evaluate the performance, a MRE is defined as $(1/N) \sum_{n=1}^{N} \|y_n - \hat{y}_n\|/\|\hat{y}_n\|$, where $n$ and $N$ are the index and the number of the testing data, respectively. Table I compares the MREs of the 1,500 regular testing data using three networks when (i) the network adopts a baseline encoder-decoder structure, in which a single encoder takes the concatenated $x_{\varepsilon_r}$ and $x_\sigma$ as a two-channel input, (ii) the bimodal encoder replaces the single encoder, while the output $F_{\varepsilon_r}$ and $F_\sigma$ are concatenated directly, and (iii) the final network includes the adaptive feature fusion module and the bimodal encoder. As shown, the MRE of the regular data achieved by the baseline network (i) is the largest, while that achieved by network (ii) is relatively smaller compared to that of network (i) due to the introduction of the bimodal encoder. The addition of the adaptive feature fusion module in network (iii) yields a further decrease in MRE, as it enhances the cross-modal connectivity and captures more informative features from $x_{\varepsilon_r}$ and $x_\sigma$. The comparative results demonstrate the effectiveness of the proposed bimodal encoder and adaptive feature fusion module. The low MRE (1.28%) of the final network (iii) shows that the predicted B-scans using the proposed method are highly close to the ones obtained by the FDTD solver. By comparing the parameters, training time, and testing time of three networks, the computational cost of network (iii) is higher than those of networks (i) and (ii). However, the computation time for obtaining one B-scan by network (iii) is only 12 milliseconds, whereas the FDTD solver requires 4.5 minutes to obtain one B-scan on the same GPU. Thereby, the speed-up achieved by network (iii) is as large as 22,500x.

To qualitatively analyze the performance, the predicted B-scans using the proposed network are compared with the ones obtained by the FDTD solver. Cases 1-6 in Fig. 3 present the comparative results for the subsurface object with various shapes, sizes, locations, permittivities, and conductivities. For the given permittivity and conductivity maps under various heterogeneous background soil distributions, the B-scans predicted by the proposed network match well with the B-scans obtained by the FDTD solver. Furthermore, the absolute errors between two sets of B-scans are too small and cannot be observed by human eyes.

TABLE I
QUALITATIVE ACCURACY AND EFFICIENCY COMPARISON

| Networks | Parameters | Training Time (h) | Testing Time (ms) | Speed-Up | MRE Regular Data | MRE Generalized Data |
|---|---|---|---|---|---|---|
| (i) | 9,436,401 | 2.94 | 8 | 33,750 | 1.55% | 4.44% |
| (ii) | 11,794,385 | 3.05 | 9 | 30,000 | 1.41% | 4.37% |
| (iii) | 12,717,521 | 3.14 | 12 | 22,500 | **1.28%** | **3.57%** |

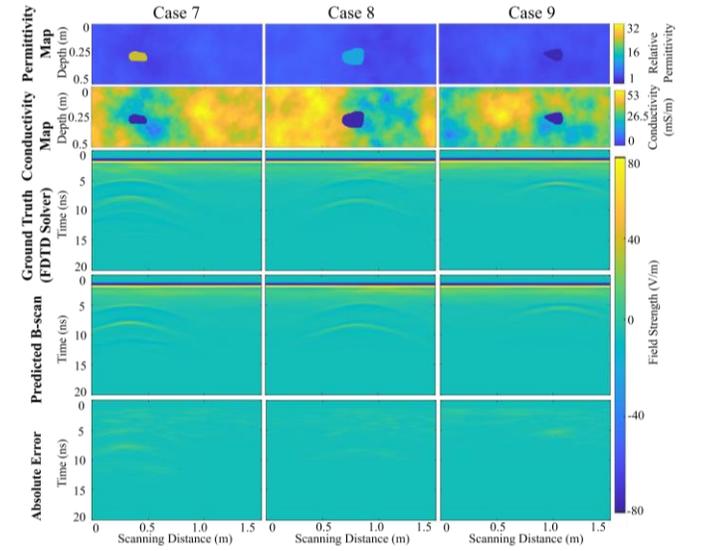

Fig. 4. The tests on the generalized data realized by hand-drawn convex objects.

### C. Generalizability Tests

#### C.1. Random Convex Objects

To test the generalization capability of the proposed network for convex objects with random shapes, three sets of permittivity and conductivity maps with hand-drawn convex objects are provided as inputs to the trained network without additional training. As shown in Table I, the network (iii) achieves the lowest MRE (3.57%) for these generalized data, which shows the improved generalization capability of the proposed network. As these hand-drawn shapes are quite different from the regular shapes in the training set, the MRE obtained for the generalized data is higher than that for the regular data but still satisfactory. The imaging results (Cases 7-9) are compared in Fig. 4. The predicted B-scans match well with the ones obtained by the FDTD solver; the differences between sets of B-scans are very small. These results demonstrate the excellent generalization capability of the proposed method for random convex objects.

#### C.2. New Scenarios

To verify the effectiveness of the proposed transfer learning algorithm, 528 sets of data (480 for training and 48 for testing) are generated for two new scenarios: (1) the random distribution of the subsurface heterogeneous background environment is replaced with a distribution different than those in the pre-training set, and (2) a commercial MALA 1.2 GHz antenna model is replaced with the previous dipole-probe configuration as the transceiver for GPR scanning. The MREs of the testing results using three models are compared in Table II. In particular, models (i), (ii), and (iii) are the pre-trained model as described in Section III.A, the model re-trained using new training data from scratch, and the model fine-tuned using transfer learning, respectively. As shown, the prediction accuracy of the model (i) is the worst, as the testing data for new scenarios are vastly different from the pre-training data, especially for the cases with the totally new antenna model. For model (ii) that performs the training from scratch using a small amount of data, the MREs are still large. This is because the network is underfitting and is far from being well-trained with a very limited training data. However, model (iii) leveraging the transfer learning technique to effectively transfer the well-learned non-linear mapping relationship in the pre-training stage to new tasks, can accurately predict the B-scans for these new scenarios. The imaging results for the new scenarios are shown in Fig. 5. The B-scans predicted using the model (iii) match well with the ones obtained by the FDTD solver.

Futher experiments have been conducted to verify the network's generalizability in new cases where the time window and the dimension of the survey area are changed. The network achieves MREs lower than 1.2% in these new cases. The quantitative and qualitative analyses in this section demonstrate that the transfer learning-applied network shows a significantly enhanced generalization capability for the new scenarios.

TABLE II
MRE COMPARISON FOR TWO NEW SCENARIOS USING THREE MODELS

| New Scenarios | Model (i) | Model (ii) | Model (iii) |
|---|---|---|---|
| (1) New Distribution | 6.88% | 4.07% | **1.20%** |
| (2) New Antenna Model | 126.10% | 12.94% | **3.91%** |

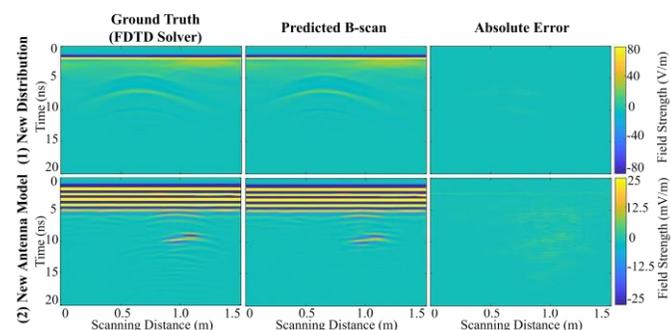

Fig. 5. The imaging results for new scenarios.

## IV. CONCLUSION

In this letter, a deep learning-based 2D GPR forward solver was proposed to obtain the B-scans for subsurface convex objects. In particular, a novel bimodal encoder-decoder neural network with the adaptive feature fusion module was designed to learn the non-linear mapping relationship between the subsurface permittivity and conductivity maps and the corresponding B-scan. Transfer learning was introduced to enhance the network's generalization capability. Given the permittivity and conductivity maps of a random convex object, the network can predict the corresponding B-scan accurately and efficiently. The computation time required by the proposed network to predict one B-scan is 22,500x less than that required by the classical FDTD solver. With its high accuracy and efficiency, the proposed network could be highly beneficial for GPR applications requiring forward solvers, such as full-wave inversion and data-driven GPR techniques. The performance of the proposed method are tested for multi-object scenarios and concave-object scenarios, for which the MREs are achieved as 5.21% and 7.50%, respectively. The performance for these cases is degraded compared with the convex-object scenarios, which are the focus of this study. Our future work will investigate advanced deep learning solvers for non-convex and multiple object scenarios.

## REFERENCES

[1] D. Goodman, "Ground-penetrating radar simulation in engineering and archaeology," *Geophysics*, vol. 59, no. 2, pp. 224-232, 1994.
[2] L. Gurel and U. Oguz, "Three-dimensional FDTD modeling of a ground-penetrating radar," *IEEE Transactions on Geoscience and Remote Sensing*, vol. 38, no. 4, pp. 1513-1521, 2000.
[3] I. Van den Bosch, S. Lambot, M. Acheroy, I. Huynen, and P. Druyts, "Accurate and efficient modeling of monostatic GPR signal of dielectric targets buried in stratified media," *Journal of Electromagnetic Waves and Applications*, vol. 20, no. 3, pp. 283-290, 2006.
[4] H. Liu, B. Xing, H. Wang, J. Cui, and B. F. Spencer, "Simulation of ground penetrating radar on dispersive media by a finite element time domain algorithm," *Journal of Applied Geophysics*, vol. 170, p. 103821, 2019.
[5] D. Feng, C. Cao, and X. Wang, "Multiscale full-waveform dual-parameter inversion based on total variation regularization to on-ground GPR data," *IEEE Transactions on Geoscience and Remote Sensing*, vol. 57, no. 11, pp. 9450-9465, 2019.
[6] Tong, Zheng, J. Gao, and D. Yuan, "Advances of deep learning applications in ground-penetrating radar: A survey," *Construction and Building Materials*, vol. 258, p. 120371, 2020.
[7] B. Liu, Y. Ren, H. Liu, H. Xu, Z. Wang, A. G. Cohn, and P. Jiang, "GPRInvNet: Deep learning-based ground-penetrating radar data inversion for tunnel linings," *IEEE Transactions on Geoscience and Remote Sensing,* 2021.
[8] T. M. Hansen, and K. S. Cordua, "Efficient Monte Carlo sampling of inverse problems using a neural network-based forward—Applied to GPR crosshole traveltime inversion," *Geophysical Journal International*, vol. 211, no. 3, pp. 1524-1533, 2017.
[9] I. Giannakis, A. Giannopoulos, and C. Warren, "A machine learning-based fast-forward solver for ground penetrating radar with application to full-waveform inversion," *IEEE Transactions on Geoscience and Remote Sensing*, vol. 57, no. 7, pp. 4417-4426, 2019.
[10] A. Vaswani, N. Shazeer, N. Parmar, J. Uszkoreit, L. Jones, A. N. Gomez, Ł. Kaiser, and I. Polosukhin, "Attention is all you need," in *Advances in Neural Information Processing Systems*, 2017, pp. 5998–6008.
[11] R. Tao, Z. Pan, R. K. Das, X. Qian, M. Z. Shou, and H. Li, " Is someone speaking? exploring long-term temporal features for audio-visual active speaker detection," in *Proceedings of the ACM International Conference on Multimedia*, 2021, pp. 3927-3935.
[12] C. Tan, F. Sun, T. Kong, W. Zhang, C. Yang, and C. Liu, "A survey on deep transfer learning," in *International Conference on Artificial Neural Networks*, 2018: Springer, pp. 270-279.
[13] C. Warren, A. Giannopoulos, and I. Giannakis, "gprMax: Open source software to simulate electromagnetic wave propagation for Ground Penetrating Radar," *Computer Physics Communications*, vol. 209, pp. 163-170, 2016.
[14] N. R. Peplinski, F. T. Ulaby, and M. C. Dobson, "Dielectric properties of soils in the 0.3-1.3-GHz range," *IEEE Transactions on Geoscience and Remote Sensing*, vol. 33, no. 3, pp. 803-807, 1995.